\documentclass[]{aa}
\usepackage{graphicx}
\usepackage[varg]{txfonts}
\usepackage{natbib}
\bibpunct{(}{)}{;}{a}{}{,} 
\usepackage[figuresright]{rotating}
\usepackage{aalongtable}
\usepackage[a4paper,breaklinks]{hyperref}
\usepackage{balance}
\idline{15}{66}

\def\teff{$T\rm_{eff}$}
\def\kms{$\mathrm{km\, s^{-1}}$}

\newcommand{\mygi}{MyGIsFOS}

\newcommand{\logg}{\ensuremath{\log g}}

\newcommand{\mlp}{\ensuremath{\alpha_{\mathrm{MLT}}}}

\newcommand{\beq}{\begin{equation}}
\newcommand{\eeq}{\end{equation}}

\begin{document}

\title{TOPoS: I. Survey design and analysis of the first sample\thanks{Based
on observations obtained at ESO Paranal Observatory,
GTO programme 189.D-0165(A)}
}

\author{
E.~Caffau\thanks{Gliese Fellow}\inst{1,2} \and
P. Bonifacio   \inst{2} \and
L. Sbordone    \inst{1,2} \and
P. Fran\c cois \inst{3,2} \and
L. Monaco      \inst{4} \and
M. Spite       \inst{2} \and
B. Plez        \inst{5} \and
R. Cayrel      \inst{2} \and
N.~Christlieb  \inst{1} \and
P. Clark       \inst{6} \and
S. Glover      \inst{6} \and
R. Klessen     \inst{6} \and
A. Koch        \inst{1} \and
H.-G. Ludwig   \inst{1,2} \and
F. Spite       \inst{2} \and
M. Steffen     \inst{7,2} \and
S.~Zaggia      \inst{8}
}

\institute{ 
Zentrum f\"ur Astronomie der Universit\"at Heidelberg, Landessternwarte, 
K\"onigstuhl 12, 69117 Heidelberg, Germany
\and
GEPI, Observatoire de Paris, CNRS, Univ. Paris Diderot, Place
Jules Janssen, 92190
Meudon, France
\and
UPJV, Universit\'e de Picardie Jules Verne, 33 Rue St Leu, F-80080 Amiens
\and
European Southern Observatory, Casilla 19001, Santiago, Chile
\and
Laboratoire Univers et Particules de Montpellier, LUPM, Universit\'e Montpellier 2, 
CNRS, 34095 Montpellier cedex 5, France
\and
Zentrum f\"ur Astronomie der Universit\"at Heidelberg,
Institut f\"ur Theoretische Astrophysik, Albert-Ueberle-Stra$\beta$e 2, 69120 Heidelberg, Germany
\and
Leibniz-Institut f\"ur Astrophysik Potsdam (AIP), An der Sternwarte 16, 14482 Potsdam, Germany
\and
Istituto Nazionale di Astrofisica,
Osservatorio Astronomico di Padova Vicolo dell'Osservatorio 5, 35122 Padova, Italy
}
\authorrunning{Caffau et al.}
\titlerunning{EMP stars}
\offprints{E.~Caffau}
\date{Received ...; Accepted ...}

\abstract%
{The metal-weak tail of the metallicity distribution function
(MDF) of the Galactic Halo stars contains crucial information
on the formation mode of the first generation of stars.
To determine this observationally, it is necessary to
observe large numbers of extremely metal-poor stars.
}
{We present here the Turn-Off Primordial Stars survey
(TOPoS) that is conducted as an ESO Large Programme at the VLT.
This project has {four} main goals:
{(i)~} to understand the formation of low-mass stars in a low-metallicity gas:
determine the metal-weak tail of the
halo MDF below [M/H]=--3.5.
In particular, we aim at determining
the critical metallicity, that is 
the lowest metallicity sufficient for the formation of
low-mass stars;
{(ii)~}to determine the relative abundance of the 
elements in extremely metal-poor stars, which are the signature of the massive first stars;
{(iii)~}to determine the trend of the 
lithium abundance at the time when the Galaxy formed; and
{(iv)~}to derive the fraction of C-enhanced extremely metal-poor
stars with respect to normal extremely metal-poor stars.
The large number of stars
observed in the SDSS provides a good sample of 
candidates of stars at extremely low metallicity.
}
{
Candidates with turn-off colours down to magnitude g=20 were selected from the low-resolution spectra
of SDSS by means of an automated procedure.
X-Shooter has the potential of performing the necessary follow-up 
spectroscopy, providing accurate metallicities 
and abundance ratios for several key elements for these stars.  
}
{
We here present the stellar parameters of the first set of stars.
The nineteen stars range in iron abundance between
--4.1 and --2.9\,dex relative to the Sun.
Two stars have a high radial velocity and, according to our estimate of their kinematics,
appear to be marginally bound to the Galaxy and are possibly accreted from another galaxy.
}
{}
\maketitle


\section{Introduction}

The $\Lambda$CDM cosmological model has recently received an impressive
confirmation from the 
high-precision measurements of the fluctuations of the cosmic 
microwave background achieved by the 
WMAP \citep{komatsu11} and Planck \citep{planck} space missions. 
From the standard Big Bang nucleosynthesis 
one can infer that three minutes after the Big Bang the baryonic matter in
the Universe was composed exclusively of isotopes of H and He and
traces of $^7$Li.  This implies that the first stellar objects to be
formed had to have this primordial composition, that is, were devoid of metals.

The WMAP data furthermore imply that at 
$z=10.4\pm 1.2$, that is around 500\,Myr after the Big Bang,
the Universe was largely reionised  \citep{komatsu11}. One possible source
of the ionising photons are the first stars, and fundamental questions are: 
{\sl How massive were they? What was the 
primordial initial mass function?} 
Possibly, all the first stars
were massive or exceedingly massive, with a very short lifetime
(see, e.g., \citealt{bromm04} and \citealt{glover05}, 
and references therein), although other more recent numerical
simulations suggest that the distribution of possible masses may have been
much broader than previously believed, and may even have been extended
to one solar mass or below (see e.g.\  \citealt{clark11,greif11}). 
If stars of mass 0.8\,M\sun
or lower were formed a few hundred Myr after the Big Bang,
they are expected to be still shining today, since their lifetime is longer than 
the Hubble time.

Even if such low-mass primordial stars do not exist, the first generation
of stars will have left behind traces that are still visible today in the form
of the metals found within the most metal-poor stars in the Galaxy. 
These extremely metal-poor stars are most likely the
oldest objects that exist in the Galaxy, 
formed within the first billion years after the Big Bang
(corresponding to a redshift of $z> 5$). Their chemical composition 
reflects the early enrichment of the matter by the first massive stars, 
which died as supernovae after reionising the Universe.

The shape of the low-mass tail of the metallicity distribution function
(MDF) of the Galactic halo population(s) provides observational clues
for understanding the mechanism(s) through which low-mass (i.e., $M
\lessapprox 1 M_{\odot}$) stars form from very low metallicity gas. If
cooling via the fine-structure lines of carbon and oxygen is needed to
facilitate low-mass star formation, a critical metallicity of
$\mathrm{[M/H]} \simeq -3.5$ has to be reached before such stars form
\citep{bromm03,frebel07}. Therefore, a vertical drop of the MDF at the
critical metallicity is expected to be detectable \citep{salvadori07}. If, on 
the other hand, dust cooling plays a dominant role, low-mass stars could
have formed from gas with an even lower metallicity \citep{schneider06,dopcke11}, and
therefore the low-metallicity tail of the MDF is expected to have a smooth shape.

Until recently, the deepest survey for searching metal-poor stars was
the Hamburg/ESO Survey (HES), which reached $V\approx 16$
\citep{christlieb08}, which increased the number of known
metal-poor stars ($[{\rm Fe/H}]<-3.0$) by about a factor of two with respect to all previous
surveys.  In terms of {\em overall} metallicity $Z$, or [M/H], the most
metal-poor stars known were the giants CD~$-38^{\circ}$~245
($\mathrm{[Fe/H]}=-4.1$), CS~22885-096, BS~16467-062, and CS~22172-002
with $\mathrm{[Fe/H]}\approx -3.8$ \citep{francois03,cayrel04}, and also
the turn-off binary star CS~22876-32 \citep{norris00,jigh08} with
$\mathrm{[Fe/H]}\approx -3.7$. That is, the total sample of stars at
$\mathrm{[M/H]} < -3.0$ or $\mathrm{[Fe/H]} < -3.0$ is still small.

\cite{schoerck09} and \cite{li10} reported that the MDF derived from the
HES, when corrected for selection biases, shows a vertical drop
approximately at $\mathrm{[Fe/H]}=-3.5$. However, the more recent study
of \cite{yong13} indicates that this drop may be an artefact caused by
the small sample size. In contrast to the aforementioned studies that were based
on HES stars alone, the low-mass tail of the MDF constructed by Yong et
al. is smooth. Even larger samples are needed to settle this question.

\section{TOPoS goals}

Turn-Off PrimOrdial Stars (TOPoS) is a survey based on the VLT@ESO 
Large Programme 189.D-0165. 
The observation programme spans four ESO semesters (89-92),
from April 2012 to March 2014, for a total of 120\,h
at X-Shooter and 30\,h at UVES. 
Seventy-five metal-poor candidates have been/will be observed with X-Shooter in total, and 
the most interesting stars, about five stars, have been/will be observed with UVES. 
The main goals of the survey TOPoS are the following:
\begin{itemize}
\item
Understand the formation of low-mass stars in a low-metallicity gas.
A fundamental question is whether do primordial low-mass stars exist.
If they do not exist, what is the value of the critical metallicity. 
\item
Constrain the masses of the first massive Pop.\,III stars
from the chemical composition of this sample of extremely metal-poor (EMP) stars.
\item
Derive the Li abundances in these EMP stars and understand the relation of these
to the primordial Li.
\item
Derive the fraction of C-enhanced extremely metal-poor
(CEMP) stars with respect to normal EMP stars.
This goal will be fulfilled for cool stars 
(\teff$<$6200\,K, for targets hotter than this limit
it is very difficult to identify a CEMP star from a low-resolution spectrum) 
from the SDSS spectra, wich are not expected to be biased in this respect. 
We tried to avoid CEMP stars in our selection. 
\end{itemize}

The TOPoS survey involves the following steps:
\begin{enumerate}
\item Select EMP candidates from SDSS (see below).
\item Follow-up observation at intermediate and high resolution.
\item 1D abundance analysis in local thermodynamical equilibrium (LTE).
\item Computation of 3D corrections and departures from LTE effects (NLTE).
\item Calibration of the low-resolution data.
\item Derivation of the MDF.
\item Kinematical properties of the stars.
\item Derive from the stellar spectra the interstellar components
to map the gas in a volume of about 5\,kpc around the Sun.
\end{enumerate}

In this paper we present the programme, describe the target selection, 
and provide the chemical analysis of Fe, Si, Mg, and Ca
based on traditional 1D model atmospheres and spectral-synthesis based
on LTE of the first sample of stars.
The effects of stellar granulation and NLTE will be analysed for the complete
sample in a dedicated paper.

We aim to use intermediate- to high-resolution (X-Shooter, UVES) spectra
to determine the metallicity of a sample of the most metal-poor 
TO stars we can select from the SDSS. This information will be used
to calibrate the huge numbers of SDSS low-resolution spectra
to gather information on the shape of the metal-poor tail of the metallicity 
distribution function of the Galactic halo at metallicities [M/H] below $-2.5$ 
and to search for the existence of a vertical drop in the MDF at $[{\rm M/H}]<-3.5$. 
With these data in hand we aim to  calibrate
the metallicity estimates obtained from the low-resolution SDSS spectra to obtain a better statistics of
the metal-poor tail of the MDF.
Our project is a spin-off of a pilot survey
conducted using French and Italian guaranteed observing  
time on  X-Shooter: a sample of twelve stars has previously been
analysed by \citet{bonifacio11,leostar,gto11,stellina}, and \citet{gto12}. 
In addition, other useful data for our purposes can be found in
\citet{sdss_uves}, who analysed 16 EMP stars observed with UVES that were
selected in a similar manner from SDSS;
\citet{spite13}, who analysed three stars selected because they were C-enhanced;
and \citet{behara}, who analysed another set of three C-enhanced stars
extracted from the SDSS.
In an effort parallel to our own, the team led by W. Aoki has observed 137 metal-poor
and extremely metal-poor stars extracted from the SDSS with HDS at the 8.2\,m SUBARU telescope
\citep{aoki}. A sizeable fraction of this sample is composed of TO stars
and will be valuable for our plan to calibrate the metallicity
estimates obtained from the SDSS spectra.

As stated above, an important goal of this survey 
is to determine the detailed chemical abundances of 
the stars with the lowest [Fe/H] in our Galaxy. 
The detailed abundance ratios of these stars convey information on
the masses of the generation of stars that produced these
metals and of which the stars we observe are the descendants.
This provides important constraints on the 
initial mass function of the first generation
of stars that enriched the Galaxy, which in turn translates
into information on the number of ionising photons provided 
by the first stars, a crucial ingredient for realistic models of
cosmological reionisation.

Another further interest in selecting a sample of TO stars is that 
the lithium abundance in their atmospheres has been shown to be
remarkably constant \citep{spite2}, irrespective of the effective temperature and
the metallicity as long as $\rm [Fe/H] \le -1.0$. 
This plateau in Li abundance (the Spite plateau) 
has a very small scatter, which has been interpreted to be the lithium abundance 
synthesised during the primordial hot and dense phase of the Universe 
(Big Bang nucleosynthesis). But its value ($\rm A(Li)\approx 2.2$)
is well below the value predicted by the baryonic
density implied, for instance, by the fluctuations of the CMB and the standard 
Big Bang nucleosynthesis model ($\rm A(Li)\approx 2.7$ \citealt{cyburt,coc}). 
Moreover, \citet{sbordone10} 
have described a meltdown of the Spite plateau below $\rm [Fe/H]\approx -2.8$.  
The more metal-poor the star is, the more often lithium appears to be depleted.
Is Li really often depleted in the most metal-poor stars
(e.g. mixing with deep layers or longer time for diffusion to operate), or is there a slope in 
the relation of the lithium abundance to metallicity? 
It is extremely important to have a larger sample of EMP TO 
stars to understand this behaviour of the lithium abundance and to try 
to derive the primordial value of the abundance of $\rm^{7}Li$,
unless some diffusion process is at work preferentially in EMP stars. 
A primordial value as low as  $\rm A(Li)\approx 2.2$ or lower 
is not compatible with the understanding of the standard Big Bang.
The \ion{Li}{i} doublet should not be detectable 
in the majority of our X-Shooter spectra because of the
low spectral resolution (see however \citealt{gto11}), but it
should be visible in the UVES spectra, if the stars have a Li abundance
compatible with the Spite plateau.

\section{Target selection}

All of the stars observed in the TOPoS survey have been
selected from stars that were observed spectroscopically
in the SDSS/SEGUE/BOSS surveys \citep[][]{york00,yanny09,dawson13}
from data releases 7, 8, and 9 \citep{dr7,dr8,dr8e,dr9}.
We selected stars based on the dereddened $(g-z)$
and $(u-g)$ colours:
$0.18 \le (g-z)_0 \le 0.70$ and $(u-g)_0 > 0.70$.
As discussed in \citet{sdss_uves}, this selects the stars
of the halo turn-off and excludes the majority
of the white dwarf stars.
We restricted the sample in magnitude by requiring $g< 20$.
We also required that the object was classified morphologically as
STAR, and attached suitable flags on the
photometric data to ensure that the object was not saturated, 
blended\footnote{not an object that ``had multiple peaks detected within it; was thus a candidate to be a deblending parent''}, 
deblended as apparently moving\footnote{not an ``object that the deblender treated as moving''}, or
contained unchecked pixels. 
We decided to take advantage of each new data release
of SDSS that became available in the course of the programme.
As soon as the data release was available, we downloaded all
the available spectra and processed them with our automatic
code Abbo\footnote{This code derives the metallicity by a $\chi^2$ fitting
of synthetic spectra to selected metallic features.} 
\citep{abbo} to obtain metallicity estimates. 
This implies that the newly released 
version of each spectrum was reprocessed each time. 
The analysis was run in parallel on our small computer cluster. 
The run analysing Data Release 9 processed
201\,181 SDSS spectra and 23\,197 BOSS spectra. 
In the course of the four observational periods, we selected 75 stars that looked the most promising
(lowest estimated metallicity) and were compatible with the observational
requirements of each period.
The temperature was fixed from the $(g-z)_0$
colour via the calibration\footnote{${\rm T}_{\rm eff}=7126.7-2844.2\left(g-z\right)+666.80\left(g-z\right)^2-11.724\left(g-z\right)^3$} 
presented in \citet{ludwig08}, and the surface gravity was fixed to
\logg$ = 4.0$ (c.g.s. units). 

When all the X-Shooter and UVES data will have been
analysed and the low-resolution estimates recalibrated, we plan
to run \mygi\ \citep{mygisfos}
in a configuration suitable for low
resolution on the latest available release of SDSS to derive
the metal-poor tail of the MDF.
To derive the MDF it is important to use a well-defined
tracer population. The population that is best adapted for this
are the turn-off stars for three reasons:    
\begin{enumerate}
\item 
the TO stars are very easy to select from colours, with only 
HB stars as a minor contaminant;
\item 
they are very numerous, and 
\item 
they are the brightest among the long-living unmixed stars:
their chemical abundances are essentially unaltered since their
formation.
\end{enumerate}
Another possible tracer population are K giants. However, selecting
K giants from colours suffers from serious contamination by K dwarfs
and the metallicity measurement can be altered by self-pollution.

\section{Observations and data reduction}

The observations were performed in service mode 
with Kueyen (VLT UT2) and the high-efficiency spectrograph 
X-Shooter \citep{dodorico,vernet}. 
The X-Shooter spectra range from 300\,nm to 2400\,nm
and are gathered by three detectors. The observations have been performed 
in staring mode with $1\times 1$  binning  and the integral field unit (IFU), 
which reimages an input field of $4\farcs{0}\times 1\farcs{8}$ 
into a pseudo-slit of $12\farcs{0}\times 0\farcs{6}$ \citep{IFU} 
As no spatial information was available 
for our targets, we used the IFU as
a slicer with three 0\farcs{6} slices. 
This corresponds to a resolving power
of R=7900 in the ultra-violet arm (UVB) and R= 12 600 in the visible arm (VIS). 
The stellar light is divided in three arms by X-Shooter; 
we analysed here only the UVB and VIS spectra.
The stars we observed are faint and have most of their flux in the blue part 
of the spectrum, so that the signal to noise ratio (S/N) of the infra-red spectra is too low 
to allow the analysis.
The spectra were reduced using the X-Shooter pipeline \citep{goldoni},
which performs the bias and background subtraction,
cosmic-ray-hit removal \citep{vandokkum01}, sky subtraction \citep{kelson03},
flat-fielding, order extraction, and merging.
However, the spectra were not reduced using the IFU pipeline
recipes. Each of the three slices of the spectra were instead reduced 
separately in slit mode with a manual localisation of the source and the sky. 
This method allowed us to perform the best possible
extraction of the spectra, leading to an efficient cleaning of 
the remaining cosmic ray hits, but also to a noticeable improvement 
in the S/N. Using the IFU can cause some problems with the sky
subtraction because there is only $\pm$ 1 arcsec on both sides of the
object. In the case of a large gradient in the spectral flux (caused
by emission lines), the modelling of the sky-background signal
can be of poor quality owing to the small number of points used
in the modelling.

\begin{table*}
\caption{\label{allstar}
Coordinates and photometric data of our programme stars.}
\tabskip=0pt
\begin{center}
\begin{tabular}{lccrrrrrr}
\hline\noalign{\smallskip}
\multicolumn{1}{l}{SDSS ID}& 
\multicolumn{1}{c}{RA}&
\multicolumn{1}{c}{Dec}& 
\multicolumn{1}{c}{$u$}& 
\multicolumn{1}{c}{$g$}& 
\multicolumn{1}{c}{$r$}&
\multicolumn{1}{c}{$i$}&
\multicolumn{1}{c}{$z$}&
\multicolumn{1}{c}{A$_{V}$}\\
 &  J2000.0 & J2000.0 & [mag] & [mag] & [mag] & [mag] &  [mag]  \\
\noalign{\smallskip}\hline\noalign{\smallskip}
J000411-055027 & 00\, 04\, 11.61 &$-$05\, 50\, 27.67 & $20.23$ & $19.30$ & $18.99$ & $18.85$ & $18.84$ & $0.04$ \\ 
J002558-101509 & 00\, 25\, 58.60 &$-$10\, 15\, 09.37 & $18.27$ & $17.35$ & $17.09$ & $17.03$ & $16.98$ & $0.05$ \\ 
J014036+234458 & 01\, 40\, 36.22 &$+$23\, 44\, 58.09 & $16.78$ & $15.82$ & $15.35$ & $15.12$ & $15.03$ & $0.12$ \\ 
J014828+150221 & 01\, 48\, 28.99 &$+$15\, 02\, 21.56 & $19.10$ & $18.25$ & $17.90$ & $17.80$ & $17.76$ & $0.05$ \\ 
J031348+011456 & 03\, 13\, 48.15 &$+$01\, 14\, 56.50 & $20.28$ & $19.35$ & $19.01$ & $18.86$ & $18.84$ & $0.09$ \\ 
J040114-051259 & 04\, 01\, 14.72 &$-$05\, 12\, 59.07 & $19.46$ & $18.58$ & $18.14$ & $17.95$ & $17.89$ & $0.09$ \\ 
J105002+242109 & 10\, 50\, 02.35 &$+$24\, 21\, 09.72 & $18.89$ & $18.04$ & $17.66$ & $17.49$ & $17.40$ & $0.02$ \\ 
J112750-072711 & 11\, 27\, 50.91 &$-$07\, 27\, 11.49 & $18.84$ & $17.92$ & $17.69$ & $17.62$ & $17.59$ & $0.04$ \\ 
J124121-021228 & 12\, 41\, 21.49 &$-$02\, 12\, 28.56 & $20.15$ & $19.31$ & $18.89$ & $18.71$ & $18.65$ & $0.03$ \\ 
J124304-081230 & 12\, 43\, 04.19 &$-$08\, 12\, 30.56 & $19.16$ & $18.25$ & $17.77$ & $17.59$ & $17.51$ & $0.03$ \\ 
J124719-034152 & 12\, 47\, 19.47 &$-$03\, 41\, 52.43 & $19.32$ & $18.50$ & $18.24$ & $18.15$ & $18.14$ & $0.03$ \\ 
J150702+005152 & 15\, 07\, 02.03 &$+$00\, 51\, 52.63 & $19.75$ & $18.77$ & $18.60$ & $18.48$ & $18.41$ & $0.06$ \\ 
J172552+274116 & 17\, 25\, 52.20 &$+$27\, 41\, 16.72 & $20.34$ & $19.36$ & $19.15$ & $19.06$ & $19.06$ & $0.05$ \\ 
J214633-003910 & 21\, 46\, 33.18 &$-$00\, 39\, 10.21 & $19.01$ & $18.14$ & $17.89$ & $17.79$ & $17.76$ & $0.06$ \\ 
J220121+010055 & 22\, 01\, 21.77 &$+$01\, 00\, 55.49 & $19.56$ & $18.65$ & $18.38$ & $18.29$ & $18.25$ & $0.06$ \\ 
J222130p000617 & 22\, 21\, 30.23 &$+$00\, 06\, 17.09 & $20.50$ & $19.51$ & $19.20$ & $18.99$ & $18.86$ & $0.07$ \\ 
J225429p062728 & 22\, 54\, 29.61 &$+$06\, 27\, 28.29 & $19.70$ & $18.81$ & $18.47$ & $18.34$ & $18.26$ & $0.08$ \\ 
J230243-094346 & 23\, 02\, 43.33 &$-$09\, 43\, 46.03 & $19.99$ & $19.11$ & $18.72$ & $18.58$ & $18.51$ & $0.04$ \\  
J235210+140140 & 23\, 52\, 10.23 &$+$14\, 01\, 40.19 & $19.10$ & $18.23$ & $17.98$ & $17.89$ & $17.84$ & $0.04$ \\    
\noalign{\smallskip}\hline\noalign{\smallskip}
\end{tabular}
\end{center}
\end{table*}

\section{Model atmospheres and spectral-synthesis}

The analysis is based on a grid of 1D plane-parallel hydrostatic model
atmospheres, computed in LTE with OSMARCS \citep{G2008}.
The grid of the model atmospheres covers the following stellar parameter ranges:
\begin{itemize}
\item effective temperature: 5200\,K$\leq$\teff$\leq$6800\,K, step of 200\,K;
\item surface gravity:  $3.5\leq$\logg$\leq 4.5$ (c.g.s. units), step of 0.5;
\item metallicity: $-2.5\leq [{\rm M/H}]\leq-4.5$, step of 0.5\,dex;
\item $\alpha$-element abundances: $-0.4\leq [\alpha/{\rm Fe}]\leq +0.8$, step of 0.4;
\item micro-turbulence of 1\,km/s.
\end{itemize}

The synthetic spectra were computed with version 12.1.1 of
{\tt turbospectrum} \citep{alvarez_plez,2012ascl.soft05004P} with a resolving power
of $\lambda/\Delta\lambda =$400\,000.
They were computed in the wavelength ranges 345-547\,nm and 845-869\,nm
for three values of micro-turbulence, 0, 1, and 2\,km/s.
The data for atomic lines are taken from VALD \citep{pisk1995,kupka1999}.
The atomic parameters for the atomic lines that can be relevant for this large programme
are an update of the data used by the First Stars survey
\citep{bonifacio09,cayrel04,francois}
and are made available here in Table\,\ref{linesdata}.

\section{Analysis}

Here we present the analysis of 19 EMP stars observed
during the first period of TOPoS, ESO period 89.
The coordinates and photometry are summarised in 
Table\,\ref{allstar}. The interstellar extinction
was derived from the \citet{schlegel} maps,
corrected as in \citet{bonifacio00}.

We derived the effective temperature from the photometry,
using the $\left(g-z\right)_0$ colour and the calibration described in \citet{ludwig08}.  
As a check for the \teff,
we fitted the wings of the H$\alpha$ lines.
The fit is based on the code described in \citet{abbo}
and uses a grid of synthetic profiles computed with a modified version of the
{\tt BALMER} code\footnote{The original version is available on-line
  at \url{http://kurucz.harvard.edu/}} on models computed 
with the ATLAS code \citep{kurucz93,SBC04,kurucz05,sbordone05}, assuming
a \mlp = 0.5 \citep{FAG93,VV96} and the opacity distribution functions of \citet{castelli03}.
This modified {\tt BALMER} code uses the theory of
\citet{barklem00,barklem00b} for self-broadening and the profiles
of \citet{stehle99} for Stark-broadening.
The \teff\ determination from the H$\alpha$ profile in EMP stars
depends on the gravity \citep{sbordone10}. A change of 0.5\,dex in gravity
can induce a change in \teff\ of up to few hundred K.
We usually prefer to rely on \teff\ from photometry, because
spectra observed with X-Shooter, which is an echelle spectrograph, 
are poorlt suited to derive the continuum placement in the H$\alpha$ region
in an objective way.

For the majority of the stars in the sample, the \teff\ derived
from the $\left(g-z\right)_0$ colour and from H$\alpha$ profile fitting
agree reasonably well (within the precision we can achieve with the S/N
of the observations and the X-Shooter spectral quality).
We found a large disagreement for one star in the sample, SDSS\,J040114-051259.
The spectrum of this star in the range of H$\alpha$ has a poor
quality, but not poor enough to justify a temperature 500\,K
cooler than the one derived from photometry. The reddening according to the 
\citet{schlegel} maps is E(B--V)=0.09, hardly compatible with the weak
interstellar \ion{Ca}{ii} K and H lines visible in the spectrum.
We adopted a \teff\ of 5500\,K, which is consistent with H$\alpha$ wings
and the $\left(g-z\right)$ colour, assuming no reddening.

The stars of the sample were selected to have typical TO colours,
therefore we assumed the typical gravity $\logg = 4.0$ for all stars.
Generally, we cannot base the gravity determination on the
equilibrium of \ion{Fe}{i} and \ion{Fe}{ii} lines.
Only few stars of the sample have \ion{Fe}{ii} lines from
which we can derive an iron abundance, and in these cases the gravity of 4.0
is consistent within the precision in the abundance determination.

The relatively low resolving power (7900) of the UVB spectra
of X-Shooter, the arm where all but the \ion{Ca}{ii} triplet lines 
can be found, does not allow one to detect weak lines,
which are necessary to derive the micro-turbulence from the relation
of abundance versus equivalent width of the lines.
We know that a change in the microturbulence parameter
of $\Delta\xi =0.5$\kms\ implies a change
in the iron abundance of $\Delta\left[{\rm Fe/H}\right]\sim 0.15$\,dex.
Calibrations are available for solar-metallicity
stars (e.g. \citealt{edvardsson93}).
As discussed in \citet{gto12} the microturbulence values
derived in the studies of \citet{sdss_uves} and \citet{sbordone10}, which
cover stars similar to the ones considered here, 
suggest that fixing the microturbulence for all stars at 1.5\kms\
is reasonable, and we followed this approach here.

\begin{figure*}
\begin{center}
\resizebox{\hsize}{!}{\includegraphics[clip=true]
{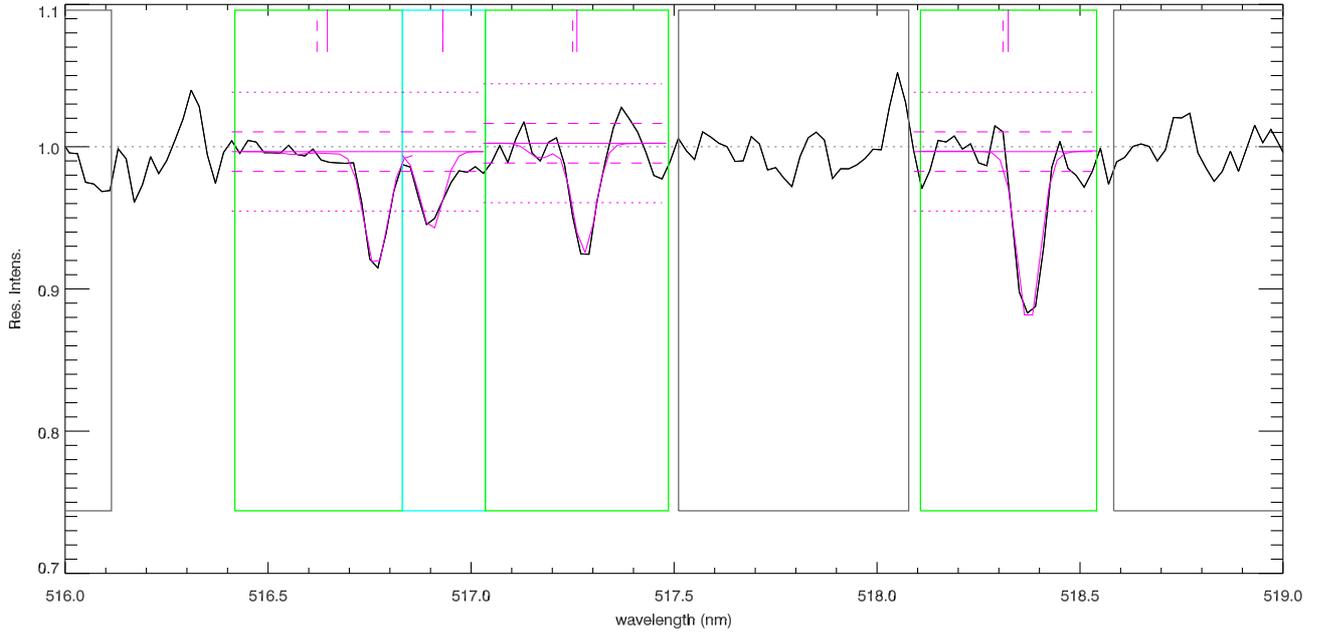}}
\end{center}
\caption[]{Observed spectrum of SDSS\,J014828+150221 (solid black) 
in the region of the \ion{Mg}{i}b triplet with the best fit superimposed (solid pink).
The green boxes enclose the range for the fit on the \ion{Mg}{i} lines, the light-blue box that for the fit on the \ion{Fe}{ii} line.
The magenta horizontal solid lines represent the continuum, the magenta horizontal dashed lines represent one $\sigma$
of the S/N ratio of the observed spectrum in the region, and the magenta horizontal dotted lines represent three 
$\sigma$ of the S/N ratio. The magenta vertical solid lines show the wavelength at the centre of the
fitting range at rest, and the magenta vertical dashed lines show the same after the shift assumed by the fitting.
}
\label{mgb}
\end{figure*}

A summary of the stellar parameters of the
sample of stars is available in Table\,\ref{analysis}.
The chemical abundances for Fe, Si, Mg, and Ca were derived with the code \mygi\ \citep{mygisfos},
based on the OSMARCS-{\tt turbospectrum} grid of synthetic spectra,
and are shown in Table\,\ref{abbondanze}.
The reference solar abundances are taken from 
\citet{abbosun} for C and Fe, and from \citet{lodders09} for the other elements.
An example of line fitting for the \ion{Mg}{i} lines performed by \mygi\ is shown in Fig.\,\ref{mgb}.

For iron, for which several lines can be detected, the $\sigma$ given in 
Table\,\ref{abbondanze} and Fig.\,\ref{plotmg} is the line-to-line scatter.
The same is true for the other elements for which more than one line can be detected.
To estimate the error
for elements for which only one line could be measured
we computed Monte Carlo simulations of 
1000 events on the \ion{Ca}{i} line at 422.8\,nm
and on the \ion{Si}{i} line at 390.5\,nm, and derived a $\sigma$ of 0.19 \citet{gto12}.
For the Monte Carlo simulation we injected a Poisson noise of 45 in a synthetic spectrum
of parameters (\teff/\logg/[Fe/H]) 6200/4.0/--3.5 for the above mentioned lines
and analysed them with \mygi.

We examined the G-band to detect CEMP stars.
We derived an upper limit from 1D synthetic spectra and found no star on the C-plateau
of \citet{spite13}.
For hot stars the upper limits are less significant because the band is weak.
We know that the molecular bands are very sensitive to granulation effects
\citep{bonifacio13}, therefore we plan to base the analysis of the G-band on 
hydrodynamical models for the complete sample of stars observed during the TOPoS survey.

The star SDSS\,J031348+011456 is on the hot side of the sample, therefore the lines appear weak
in the spectrum, even though this is not one of the most metal-poor objects of the sample.
The low S/N induces a high line-to-line scatter in the abundances.

For star SDSS\,J214633-003910,
the \ion{Mg}{i} line at 470.29\,nm gives an abundance about 2\,dex
higher than the abundance derived from the other Mg lines. 
We have no explanation for this, but we removed this line from the analysis.

Typically, the calcium abundance derived from the \ion{Ca}{ii}
lines of the IR triplet is very high, on average by 0.4\,dex higher
than the abundance obtained from the \ion{Ca}{i} lines.
We are aware that the sky subtraction, when observing with the IFU,
is difficult in the visible arm. In addition, the Ca triplet lines are affected
by NLTE \citep{mashonkina} and granulation effects.
\citet{spite12} derived NLTE corrections for the Ca triplet lines of about --0.4\,dex
for an effective temperature in the range of 6000-6500\,K, a gravity of 4, and $[{\rm Fe/H}]=-3.5$.
The specific problems affecting these lines will be addressed when the complete
sample of stars is available.

\begin{table*}
\caption{\label{analysis}
Stellar parameters and main results.
}
\begin{center}
\begin{tabular}{lrrrrrrr}
\hline\noalign{\smallskip}
Star                 &  ${\rm T_{\rm eff}}$ & \logg & $\xi$ &S/N & [Fe/H]$_{\rm SDSS}$ & [Fe/H] & A(C) \\ 
                    &   & K                 & \kms  &  @ 400\,nm &      &    \\
\hline\noalign{\smallskip}
 SDSS\,J000411-055027 & 6174 & 4.0 & 1.5 &  30 & $-4.44$ & $-2.96$ & $\leq 6.5$ \\ 
 SDSS\,J002558-101509 & 6408 & 4.0 & 1.5 &  40 & $-3.53$ & $-3.08$ & $\leq 6.5$ \\ 
 SDSS\,J014036+234458 & 5848 & 4.0 & 1.5 &  58 & $-3.65$ & $-3.83$ & $\leq 6.0$ \\ 
 SDSS\,J014828+150221 & 6151 & 4.0 & 1.5 &  52 & $-3.11$ & $-3.41$ & $\leq 6.0$ \\ 
 SDSS\,J031348+011456 & 6335 & 4.0 & 1.5 &  29 & $-3.15$ & $-3.31$ & $\leq 7.0$ \\ 
 SDSS\,J040114-051259 & 5500 & 4.0 & 1.5 &  29 & $-3.04$ & $-3.62$ & $\leq 5.0$ \\ 
 SDSS\,J105002+242109 & 5682 & 4.0 & 1.5 &  40 & $-3.16$ & $-3.93$ & $\leq 6.2$ \\ 
 SDSS\,J112750-072711 & 6474 & 4.0 & 1.5 &  67 & $-3.30$ & $-3.34$ & $< 7.0$    \\ 
 SDSS\,J124121-021228 & 5672 & 4.0 & 1.5 &  57 & $-3.04$ & $-3.47$ & $\leq 6.0$ \\ 
 SDSS\,J124304-081230 & 5488 & 4.0 & 1.5 &  40 & $-3.17$ & $-3.92$ & $\leq 5.5$ \\ 
 SDSS\,J124719-034152 & 6332 & 4.0 & 1.5 &  73 & $-3.41$ & $-4.11$ & $\leq 6.0$ \\ 
 SDSS\,J150702+005152 & 6555 & 4.0 & 1.5 &  40 & $-4.46$ & $-3.51$ & $\leq 7.0$ \\ 
 SDSS\,J172552+274116 & 6624 & 4.0 & 1.5 &  30 & $-3.22$ & $-2.91$ & $\leq 7.0$ \\ 
 SDSS\,J214633-003910 & 6475 & 4.0 & 1.5 &  64 & $-3.12$ & $-3.14$ & $\leq 6.5$ \\ 
 SDSS\,J220121+010055 & 6392 & 4.0 & 1.5 &  49 & $-3.54$ & $-3.03$ & $\leq 6.5$ \\ 
 SDSS\,J222130+000617 & 5891 & 4.0 & 1.5 &  25 & $-3.48$ & $-3.14$ & $\leq 6.0$ \\ 
 SDSS\,J225429+062728 & 6169 & 4.0 & 1.5 &  26 & $-3.52$ & $-3.01$ & $\leq 5.5$ \\ 
 SDSS\,J230243-094346 & 5861 & 4.0 & 1.5 &  43 & $-5.55$ & $-3.71$ & $\leq 5.5$ \\ 
 SDSS\,J235210+140140 & 6313 & 4.0 & 1.5 &  54 & $-3.30$ & $-3.54$ & $\leq 6.0$ \\ 
\noalign{\smallskip}\hline\noalign{\smallskip}
\end{tabular}
\end{center}
\end{table*}

\begin{table*}
\setlength{\tabcolsep}{2pt}
\caption{Abundances \label{abbondanze}}
\centering
\begin{tabular}{lcccccccccccccccccr}
\hline\hline
Star        &[\ion{Fe}{i}/H]&  $\sigma$ & N &[\ion{Fe}{ii}/H]&  $\sigma$ & N & [Mg/H] & $\sigma$ & N &  [Si/H]&  N & [\ion{Ca}{i}/H]  & $\sigma$ & N &
  [\ion{Ca}{ii}/H] & $\sigma$ & N \\ 
\hline
\object{SDSS\,J000411-055027} & $-2.96$ & 0.32 & 13 &         &      &   & $-2.42$ & 0.38 & 5 &         &   & $-3.33$ &     & 1 & $-2.43$ & 0.13 & 2 \\
\object{SDSS\,J002558-101509} & $-3.08$ & 0.23 & 14 &         &      &   & $-2.89$ & 0.18 & 5 &         &   & $-3.20$ &     & 1 & $-2.02$ & 0.31 & 3 \\
\object{SDSS\,J014036+234458} & $-3.83$ & 0.21 & 27 &         &      &   & $-3.38$ & 0.21 & 4 & $-3.82$ & 1 & $-3.78$ &     & 1 & $-3.20$ & 0.13 & 3 \\
\object{SDSS\,J014828+150221} & $-3.41$ & 0.22 & 25 & $-3.17$ &      & 1 & $-3.26$ & 0.18 & 3 &         &   & $-3.88$ &     & 1 & $-2.71$ & 0.34 & 2 \\
\object{SDSS\,J031348+011456} & $-3.31$ & 0.42 &  6 &         &      &   &         &      &   &         &   & $-3.10$ &     & 1 & $-2.79$ &      & 1 \\
\object{SDSS\,J040114-051259} & $-3.69$ & 0.36 &  7 &         &      &   & $-3.56$ & 0.20 & 3 & $-3.42$ & 1 & $-3.18$ &     & 1 &         &      &   \\
\object{SDSS\,J105002+242109} & $-3.93$ & 0.27 & 11 & $-3.49$ &      & 1 & $-2.99$ & 0.05 & 3 & $-3.43$ & 1 & $-3.98$ &     & 1 & $-2.95$ & 0.09 & 2 \\
\object{SDSS\,J112750-072711} & $-3.34$ & 0.30 & 17 &         &      &   & $-3.13$ & 0.09 & 3 &         &   & $-3.38$ &     & 1 & $-1.94$ &      & 1  \\
\object{SDSS\,J124121-021228} & $-3.47$ & 0.22 & 40 & $-3.39$ &      & 1 & $-2.95$ & 0.11 & 5 & $-3.35$ & 1 & $-3.99$ &     & 1 & $-2.79$ &      & 1  \\
\object{SDSS\,J124304-081230} & $-3.92$ & 0.28 & 24 &         &      &   & $-3.40$ & 0.23 & 5 & $-3.36$ & 1 & $-3.73$ &     & 1 & $-2.95$ & 0.04 & 3 \\
\object{SDSS\,J124719-034152} & $-4.11$ & 0.18 & 11 &         &      &   & $-3.55$ & 0.30 & 3 &         &   & $-3.96$ &     & 1 & $-2.78$ & 0.13 & 2 \\
\object{SDSS\,J150702+005152} & $-3.51$ & 0.17 &  7 &         &      &   & $-3.05$ & 0.21 & 4 &         &   &         &     &   & $-2.33$ & 0.11 & 3 \\
\object{SDSS\,J172552+274116} & $-2.91$ & 0.38 &  6 &         &      &   & $-2.89$ & 0.20 & 4 &         &   & $-3.29$ &     & 1 & $-2.31$ & 0.14 & 3 \\
\object{SDSS\,J214633-003910} & $-3.14$ & 0.28 & 30 & $-3.24$ &      & 1 & $-2.91$ & 0.20 & 4 & $-3.56$ & 1 & $-3.11$ &     & 1 & $-2.47$ & 0.29 & 3 \\
\object{SDSS\,J220121+010055} & $-3.03$ & 0.23 & 23 & $-3.04$ & 0.15 & 2 & $-2.62$ & 0.17 & 5 & $-3.77$ & 1 & $-3.25$ &     & 1 & $-2.00$ & 0.21 & 3 \\
\object{SDSS\,J222130+000617} & $-3.14$ & 0.15 & 10 &         &      &   & $-3.17$ & 0.14 & 2 &         &   &         &     &   & $-2.83$ & 0.20 & 2 \\
\object{SDSS\,J225429+062728} & $-3.01$ & 0.30 & 10 &         &      &   & $-2.54$ & 0.36 & 2 &         &   & $-3.28$ &     & 1 & $-2.05$ & 0.03 & 3 \\
\object{SDSS\,J230243-094346} & $-3.71$ & 0.27 & 14 &         &      &   & $-3.32$ & 0.19 & 5 & $-3.61$ & 1 &         &     &   & $-3.47$ & 0.14 & 2 \\
\object{SDSS\,J235210+140140} & $-3.54$ & 0.26 & 13 &         &      &   & $-2.94$ & 0.25 & 4 & $-3.57$ & 1 &         &     &   & $-2.16$ & 0.14 & 3 \\
\hline
\end{tabular}
\end{table*}

\section{Kinematics}

We performed a preliminar analysis of the kinematic properties of the two 
stars with the most extreme radial velocities. A more extended evaluation is
deferred to a forthcoming paper on the kinematics of the TOPOS stars. The stars 
are SDSS\,J112750-072711 and SDSS\,J014828+150221, which have a radial velocity of $419$\kms\ and $-411$\kms, respectively, with 
an uncertainty of $\simeq 10$\kms. 

We calculated the orbital 
solutions using a standard \citet{allen91} model, after obtaining the distance of each star by fitting the SDSS photometry with 
a set of metal-poor 13\,Gyr old isochrones from \citet{girardi04}. Since the colours
of the stars are compatible with either main sequence (MS), sub-giant branch (SGB), or horizontal branch (HB) stars,
we adopted the distance for the MS star.
The distance from the Sun of both stars is approximately
$\sim 4$\,kpc. For both stars we have at our disposal two estimates for the 
proper motion: one from the SDSS\,DR9, and one from PPMXL \citep{roeser10}. Although the values 
for each star are compatible within errors (the reference datasets are the same), 
we decided to explore all possible 
orbital solutions on a grid of $-40\div+40$\,mas/yr in both RA and DEC, keeping 
${\rm v}_{\rm rad}$ and distance fixed. This allowed us to reach a more reliable conclusion 
on the real solution given the high uncertainties of the proper motions.
For both stars we computed the eccentricity of the orbit starting from the
orbital solution, not from the proper motion.
This computation provides a better intrinsic shape of the orbit
than the kinetatic value.

We found that star  SDSS\,J112750-072711 is on a highly radial orbit with an eccentricity of $\sim0.97$ and
a minimum radial distance of $\sim 6$~kpc: the star is actually approaching its 
minimum orbital radial distance. The maximum radial orbital distance varies from a lower value of 
$\sim85$\,kpc for a zero value of the proper motion to values higher than 200\,kpc
in the case of proper motions larger than $\sim10$\,mas/yr, as are indeed measured. We can consider a star
with such a large maximum radial orbital distance a real outer halo object, loosely 
connected with the Milky Way and possibly belonging to some external galaxy.

The conclusions are similar for star  
SDSS\,J014828+150221. The orbit is quite radial with an
eccentricity of $\sim0.90$ and a minimum radial distance of
$\sim 6$~kpc. ${\rm R}_{\rm MAX}$ varies from $80$\,kpc
to more than 200 for a proper motion of $\sim 12$\,mas/yr. Considering that for this star 
the estimated proper motion is $\sim4$\,mas/yr, it is more probable that this star is
an outer halo star.

The extreme radial velocity is a clear signature that these stare are outer halo 
members. If the high proper motion is confirmed for SDSS\,J112750-072711 we can
consider this as an extragalactic star at its first visit to the Milky Way.

\section{Discussion}

\begin{figure}
\begin{center}
\resizebox{\hsize}{!}{\includegraphics[clip=true]
{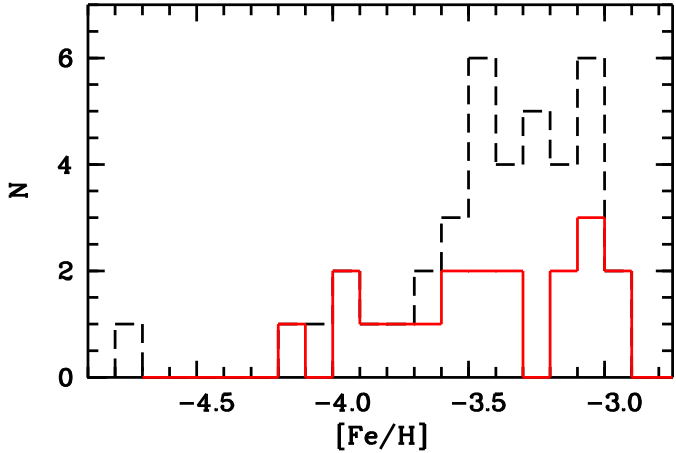}}
\end{center}
\caption[]{The histogram of the metallicities for this sample of stars (solid red)
compared to the one (1D-LTE [Fe/H] value) from our complete sample of stars (dashed red) 
selected from SDSS and observed with X-Shooter \citep{gto11,stellina} and UVES \citep{sdss_uves}.
}
\label{histmet}
\end{figure}

We analysed a sub-sample of EMP stars observed during the Large Programme TOPoS.
The selection was set with the aim to find the most metal-poor TO stars, but
we also selected relatively bright stars for X-Shooter
to calibrate the Galactic MDF.
For this sample of 19 stars, [Fe/H] ranges from $-4.1$ to $-2.9$ and the average value is $-3.32$.
In Fig.\,\ref{histmet} we show the metallicity histogram for this sample of stars.
We found no evidence for a drop at [Fe/H]=--3.5 in contrast to the results of \citet{schoerck09},
although we note that this result was already expected from the figure\,3 of \citet{ludwig08}.
There is a disagreement between [Fe/H] we derive from X-Shooter spectra
and the metallicity we derive from SDSS spectra of up to about 2\,dex,
as can also be seen from figure\,5 of \citet{gto11}, especially for the lowest metallicity.
We recall that the metallicity derived from SDSS spectra is often based on the lines of
$\alpha$ elements (e.g. the \ion{Mg}{i}b triplet) and in some cases
only on the \ion{Ca}{ii}-K line, which can be contaminated by a substantial
contribution by the inter-stellar component.
The analysis of X-Shooter spectra allows us to derive [Fe/H], [$\alpha$/Fe], and usually
also a useful upper limit on the carbon abundance.

\begin{figure}
\begin{center}
\resizebox{\hsize}{!}{\includegraphics[clip=true]
{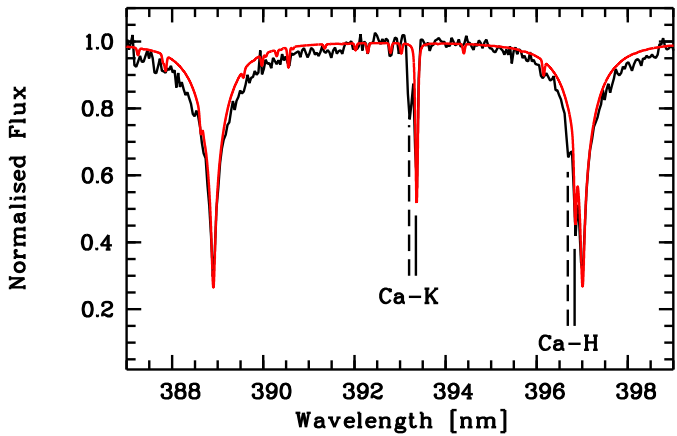}}
\end{center}
\caption[]{Spectrum of SDSS\,J124719-034152 in the range of the \ion{Ca}{ii}-H and -K lines
(solid black) compared with a synthetic spectrum of metallicity $-4$ (solid red).
The Ca stellar lines are marked with a solid black line, the interstellar
components with a dashed black line.
}
\label{cakj124719}
\end{figure}

Of the 19 stars analysed, one has $[{\rm Fe/H}] <-4$ (SDSS\,J124719-034152). 
In Fig.\,\ref{cakj124719} its wavelength range around the \ion{Ca}{ii} H and K lines is shown.
For seven stars $-4<[{\rm Fe/H}]<-3.5$.
From the present small sample it is clear that there is no cut-off
in the MDF at $[{\rm Fe/H}]\approx -3.5$, because eight stars lie below this value.
We can confirm the low metallicity of all the stars, but we have not yet found
another extreme object like SDSS\,J102715+172927 \citep{leostar}.
SDSS\,J102715+172927 was selected with five other objects and happened to
be an outlier. Statistically, we could expect another star of this type
in a sample of 19 stars. However, it can be misleading to assamble statistics from such small
numbers, but it may be that EMP stars not enhanced in carbon, with $[{\rm Fe/H}]<-4.5$,
are extremely rare, and much lower in number than CEMP stars with the same limit in iron.

\begin{figure*}
\begin{center}
\resizebox{\hsize}{!}{\includegraphics[clip=true]
{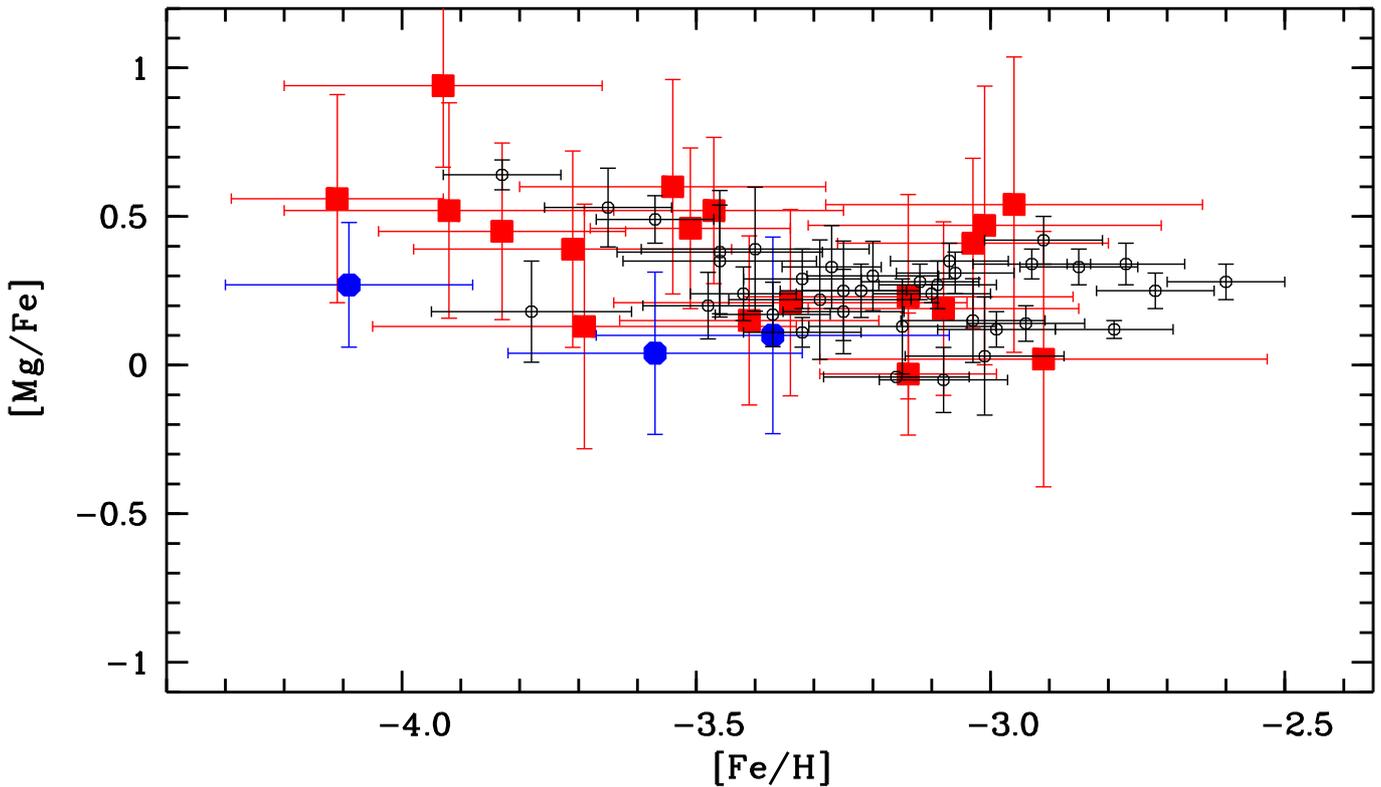}}
\end{center}
\caption[]{[Mg/Fe] vs. [Fe/H] diagram for the programme stars (filled red squares)
compared with the three stars analysed in \citet{gto12}
(filled blue hexagons)
and those in the samples of \citet{bonifacio09}
and \citet{sdss_uves} (black open circle). 
}
\label{plotmg}
\end{figure*}

The most suitable element for investigating in the
$\alpha$-enhancement in our sample of stars is Mg, because the Mgb
triplet falls in the UVB range as well as some other \ion{Mg}{i} lines.
In Fig.\,\ref{plotmg}, [Mg/Fe] vs. [Fe/H] 
is compared with that of other EMP stars.
As stated in \citet{gto12} (filled blue circles, in Fig.\ref{plotmg}),
where we discussed two EMP stars low in $\alpha$,
[Mg/Fe] shows a large scatter and several stars show a low value.
The two halo objects with high radial velocity discussed in Sec.\,7
are not extremely low in [Mg/Fe] (+0.15 and +0.21 for SDSS\,J014828+150221
and SDSS\,J112750-072711, respectively).
A more reliable statement will be possible with the complete sample
and after appropriate NLTE and 3D computations.
It is possible that a part of the scatter seen in the present
analysis is due to the neglect of 3D and NLTE effects.

\begin{acknowledgements}
EC, LS, NC, PC, SG, RK, AK, and HGL acknowledge financial support
by the Sonderforschungsbereich SFB881 ``The Milky Way
System'' (subprojects A4 and A5) of the German Research Foundation
(DFG).
We acknowledge support from the Programme National
de Physique Stellaire (PNPS) and the Programme National
de Cosmologie et Galaxies (PNCG) of the Institut National des Sciences
de l'Univers of CNRS.
P.C., S.G., and R.K. furthermore acknowledges the support of the Landesstiftung Baden-W\"{u}rttemberg 
under research contract P-LS-SPll/18 (in the program {\em Internationale Spitzenforschung ll}).

\end{acknowledgements}


%
%
%


\balance
\bibliographystyle{aa}

\Online

{\scriptsize
\begin{longtable}{lrrr}
\caption{Atomic lines.}\\
\hline\hline
\label{linesdata}
 Element & $\lambda$   & ${\rm E}_{\rm low}$ & log\,gf \\
         & [nm]        &  [eV]  & \\
\hline
\endfirsthead
\caption{continued.}\\
\hline\hline
 Element & $\lambda$   & ${\rm E}_{\rm low}$ & log\,gf \\
         & [nm]        &  [eV]  & \\
\hline
\endhead
\hline
\endfoot
\ion{Mg}{i}  & 382.9355 & 2.709 & $-0.231$ \\ 
\ion{Mg}{i}  & 383.2299 & 2.712 & $-0.356$ \\ 
\ion{Mg}{i}  & 383.2304 & 2.712 & $+0.021$ \\ 
\ion{Mg}{i}  & 383.8292 & 2.717 & $+0.392$ \\ 
\ion{Mg}{i}  & 383.8295 & 2.717 & $-0.335$ \\ 
\ion{Mg}{i}  & 383.8290 & 2.717 & $-1.530$ \\ 
\ion{Mg}{i}  & 470.2991 & 4.346 & $-0.666$ \\
\ion{Mg}{i}  & 516.7321 & 2.709 & $-0.931$ \\ 
\ion{Mg}{i}  & 517.2684 & 2.712 & $-0.450$ \\ 
\ion{Mg}{i}  & 518.3604 & 2.717 & $-0.239$ \\ 
\ion{Si}{i}  & 390.5523 & 1.909 & $-0.743$ \\ 
\ion{Ca}{i}  & 422.6728 & 0.000 & $+0.265$ \\ 
\ion{Ca}{i}  & 443.4957 & 1.886 & $-0.007$ \\ 
\ion{Ca}{i}  & 445.4779 & 1.899 & $+0.335$ \\ 
\ion{Ca}{ii} & 393.3663 & 0.000 & $+0.105$ \\ 
\ion{Ca}{ii} & 849.8023 & 1.692 & $-1.469$ \\ 
\ion{Ca}{ii} & 854.2091 & 1.700 & $-0.514$ \\ 
\ion{Ca}{ii} & 866.2141 & 1.692 & $-0.770$ \\ 
\ion{Fe}{i}  & 355.4924 & 2.832 & $+0.538$ \\ 
\ion{Fe}{i}  & 355.6878 & 2.851 & $-0.040$ \\ 
\ion{Fe}{i}  & 355.8515 & 0.990 & $-0.629$ \\ 
\ion{Fe}{i}  & 356.5379 & 0.958 & $-0.133$ \\ 
\ion{Fe}{i}  & 356.5582 & 2.865 & $-0.496$ \\ 
\ion{Fe}{i}  & 357.0098 & 0.915 & $+0.153$ \\ 
\ion{Fe}{i}  & 357.0254 & 2.808 & $+0.728$ \\ 
\ion{Fe}{i}  & 358.1193 & 0.859 & $+0.406$ \\ 
\ion{Fe}{i}  & 360.8859 & 1.011 & $-0.100$ \\ 
\ion{Fe}{i}  & 364.7843 & 0.915 & $-0.194$ \\ 
\ion{Fe}{i}  & 370.7822 & 0.087 & $-2.407$ \\ 
\ion{Fe}{i}  & 370.7920 & 2.176 & $-0.466$ \\ 
\ion{Fe}{i}  & 370.9246 & 0.915 & $-0.646$ \\ 
\ion{Fe}{i}  & 371.9935 & 0.000 & $-0.431$ \\ 
\ion{Fe}{i}  & 372.7619 & 0.958 & $-0.631$ \\ 
\ion{Fe}{i}  & 374.3362 & 0.990 & $-0.785$ \\ 
\ion{Fe}{i}  & 374.3468 & 3.573 & $+0.146$ \\ 
\ion{Fe}{i}  & 374.5561 & 0.087 & $-0.771$ \\ 
\ion{Fe}{i}  & 374.5899 & 0.121 & $-1.335$ \\ 
\ion{Fe}{i}  & 375.8233 & 0.958 & $-0.027$ \\ 
\ion{Fe}{i}  & 376.3789 & 0.990 & $-0.238$ \\ 
\ion{Fe}{i}  & 376.5539 & 3.237 & $+0.482$ \\ 
\ion{Fe}{i}  & 376.7192 & 1.011 & $-0.390$ \\ 
\ion{Fe}{i}  & 378.7880 & 1.011 & $-0.860$ \\ 
\ion{Fe}{i}  & 379.0093 & 0.990 & $-1.761$ \\ 
\ion{Fe}{i}  & 379.5002 & 0.990 & $-0.761$ \\ 
\ion{Fe}{i}  & 381.2964 & 0.958 & $-1.047$ \\ 
\ion{Fe}{i}  & 381.5840 & 1.485 & $+0.237$ \\ 
\ion{Fe}{i}  & 382.0425 & 0.859 & $+0.119$ \\ 
\ion{Fe}{i}  & 382.4444 & 0.000 & $-1.362$ \\ 
\ion{Fe}{i}  & 382.5881 & 0.915 & $-0.037$ \\ 
\ion{Fe}{i}  & 382.7822 & 1.557 & $+0.062$ \\ 
\ion{Fe}{i}  & 384.0437 & 0.990 & $-0.506$ \\ 
\ion{Fe}{i}  & 384.1048 & 1.608 & $-0.045$ \\ 
\ion{Fe}{i}  & 384.9966 & 1.011 & $-0.871$ \\ 
\ion{Fe}{i}  & 385.0818 & 0.990 & $-1.734$ \\ 
\ion{Fe}{i}  & 385.6371 & 0.052 & $-1.286$ \\ 
\ion{Fe}{i}  & 385.9911 & 0.000 & $-0.710$ \\ 
\ion{Fe}{i}  & 386.5523 & 1.011 & $-0.982$ \\ 
\ion{Fe}{i}  & 387.2501 & 0.990 & $-0.928$ \\ 
\ion{Fe}{i}  & 387.8573 & 0.087 & $-1.379$ \\ 
\ion{Fe}{i}  & 387.8018 & 0.958 & $-0.914$ \\ 
\ion{Fe}{i}  & 388.6282 & 0.052 & $-1.076$ \\ 
\ion{Fe}{i}  & 388.7048 & 0.915 & $-1.144$ \\ 
\ion{Fe}{i}  & 389.5656 & 0.110 & $-1.670$ \\ 
\ion{Fe}{i}  & 389.8009 & 1.011 & $-2.018$ \\ 
\ion{Fe}{i}  & 389.7890 & 2.692 & $-0.736$ \\ 
\ion{Fe}{i}  & 389.9707 & 0.087 & $-1.531$ \\ 
\ion{Fe}{i}  & 390.2945 & 1.557 & $-0.466$ \\ 
\ion{Fe}{i}  & 390.6479 & 0.110 & $-2.243$ \\ 
\ion{Fe}{i}  & 392.0258 & 0.121 & $-1.746$ \\ 
\ion{Fe}{i}  & 392.2912 & 0.052 & $-1.651$ \\ 
\ion{Fe}{i}  & 392.7920 & 0.110 & $-1.522$ \\ 
\ion{Fe}{i}  & 393.0297 & 0.087 & $-1.491$ \\ 
\ion{Fe}{i}  & 400.5242 & 1.557 & $-0.610$ \\ 
\ion{Fe}{i}  & 404.5812 & 1.485 & $+0.280$ \\ 
\ion{Fe}{i}  & 406.2441 & 2.845 & $-0.862$ \\ 
\ion{Fe}{i}  & 407.1738 & 1.608 & $-0.022$ \\ 
\ion{Fe}{i}  & 413.2058 & 1.608 & $-0.675$ \\ 
\ion{Fe}{i}  & 414.3868 & 1.557 & $-0.511$ \\ 
\ion{Fe}{i}  & 414.3415 & 3.047 & $-0.204$ \\ 
\ion{Fe}{i}  & 418.1755 & 2.831 & $-0.371$ \\ 
\ion{Fe}{i}  & 418.7795 & 2.425 & $-0.554$ \\ 
\ion{Fe}{i}  & 418.7039 & 2.449 & $-0.548$ \\ 
\ion{Fe}{i}  & 420.2029 & 1.485 & $-0.708$ \\ 
\ion{Fe}{i}  & 423.5936 & 2.425 & $-0.341$ \\ 
\ion{Fe}{i}  & 425.0787 & 1.557 & $-0.714$ \\ 
\ion{Fe}{i}  & 425.0119 & 2.469 & $-0.405$ \\ 
\ion{Fe}{i}  & 426.0474 & 2.399 & $-0.109$ \\ 
\ion{Fe}{i}  & 427.1760 & 1.485 & $-0.164$ \\ 
\ion{Fe}{i}  & 427.1153 & 2.449 & $-0.349$ \\ 
\ion{Fe}{i}  & 428.2403 & 2.176 & $-0.779$ \\ 
\ion{Fe}{i}  & 430.7902 & 1.557 & $-0.073$ \\ 
\ion{Fe}{i}  & 432.5762 & 1.608 & $+0.006$ \\ 
\ion{Fe}{i}  & 437.5930 & 0.000 & $-3.031$ \\ 
\ion{Fe}{i}  & 438.3545 & 1.485 & $+0.200$ \\ 
\ion{Fe}{i}  & 440.4750 & 1.557 & $-0.142$ \\ 
\ion{Fe}{i}  & 441.5122 & 1.608 & $-0.615$ \\ 
\ion{Fe}{i}  & 442.7310 & 0.052 & $-2.924$ \\ 
\ion{Fe}{i}  & 446.1653 & 0.087 & $-3.210$ \\ 
\ion{Fe}{i}  & 446.6551 & 2.831 & $-0.600$ \\ 
\ion{Fe}{i}  & 447.6019 & 2.845 & $-0.819$ \\ 
\ion{Fe}{i}  & 448.2170 & 0.110 & $-3.501$ \\ 
\ion{Fe}{i}  & 449.4563 & 2.198 & $-1.136$ \\ 
\ion{Fe}{i}  & 452.8614 & 2.176 & $-0.822$ \\ 
\ion{Fe}{i}  & 489.1492 & 2.851 & $-0.111$ \\ 
\ion{Fe}{i}  & 489.0755 & 2.875 & $-0.394$ \\ 
\ion{Fe}{i}  & 492.0502 & 2.832 & $+0.068$ \\ 
\ion{Fe}{i}  & 491.8954 & 4.154 & $-0.672$ \\ 
\ion{Fe}{i}  & 491.8994 & 2.865 & $-0.342$ \\ 
\ion{Fe}{i}  & 495.7596 & 2.808 & $+0.233$ \\ 
\ion{Fe}{i}  & 495.7298 & 2.851 & $-0.407$ \\ 
\ion{Fe}{i}  & 495.7682 & 4.191 & $-0.400$ \\ 
\ion{Fe}{i}  & 501.2068 & 0.859 & $-2.642$ \\ 
\ion{Fe}{i}  & 522.7189 & 1.557 & $-1.228$ \\ 
\ion{Fe}{i}  & 526.9537 & 0.859 & $-1.327$ \\ 
\ion{Fe}{i}  & 527.0356 & 1.608 & $-1.339$ \\ 
\ion{Fe}{i}  & 532.8039 & 0.915 & $-1.466$ \\ 
\ion{Fe}{i}  & 532.8531 & 1.557 & $-1.850$ \\ 
\ion{Fe}{i}  & 537.1489 & 0.958 & $-1.645$ \\ 
\ion{Fe}{i}  & 542.9696 & 0.958 & $-1.879$ \\ 
\ion{Fe}{ii} & 492.3927 & 2.891 & $-1.504$ \\ 
\ion{Fe}{ii} & 501.8440 & 2.891 & $-1.100$ \\ 
\ion{Fe}{ii} & 516.9033 & 2.891 & $-1.000$ \\ 
\hline                                                                   
\\
\end{longtable}
}

\end{document}